\documentclass[conference,twocolumn]{IEEEtran}
\usepackage[]{times}
\usepackage{epsfig}
\usepackage{subfigure}
\usepackage{amsmath}
\usepackage{amssymb}
\usepackage{graphicx}
\usepackage{multirow}
\usepackage{algorithm2e}
\usepackage{mathrsfs}
\usepackage{url}
\usepackage[usenames]{color}
\newcommand{\sgn}{\text{sgn}}

\newtheorem{defn} {Definition}
\newtheorem{cor} {Corollary}
\newtheorem{te}{Theorem}

\newtheorem{ex}{Example}

\newcommand{\supp}{\mbox{supp}}
\newcommand{\mrm}{\mathrm}

\begin{document}
\title{Multilevel Decoders Surpassing Belief Propagation on the Binary Symmetric Channel}
\author{\IEEEauthorblockN{Shiva Kumar Planjery}
\IEEEauthorblockA{Dept. of Electrical and Computer Eng.\\
University of Arizona\\
Tucson, AZ 85721, U.S.A.\\
Email: shivap@ece.arizona.edu}
\and
\IEEEauthorblockN{David Declercq}
\IEEEauthorblockA{ETIS\\
ENSEA/UCP/CNRS UMR 8051\\
95014 Cergy-Pontoise, France\\
Email: declercq@ensea.fr}
\and
\IEEEauthorblockN{Shashi Kiran Chilappagari, Bane Vasi$\acute{\mathrm{c}}$}
\IEEEauthorblockA{Dept. of Electrical and Computer Eng.\\
University of Arizona\\
Tucson, AZ 85721, U.S.A.\\
Email: \{shashic,vasic\}@ece.arizona.edu}
}

\maketitle
\vspace{-0.5in}
\begin{abstract}
In this paper, we propose a new class of quantized message-passing decoders for LDPC codes over the BSC. The messages take values (or levels) from a finite set. The update rules do not mimic belief propagation but instead are derived using the knowledge of  trapping sets. We show that the update rules can be derived to correct certain error patterns that are uncorrectable by algorithms such as BP and min-sum. In some cases even with a small message set, these decoders can guarantee correction of a higher number of errors than BP and min-sum.  We provide particularly good 3-bit decoders for 3-left-regular LDPC codes. They significantly outperform the BP and min-sum decoders, but more importantly, they achieve this at only a fraction of the complexity of the BP and min-sum decoders.
\end{abstract}
%
\section{Introduction}\label{sect_Intro}
Low-density parity-check (LDPC) \cite{gallager} codes have received much attention in the past several years owing to their exceptional performance under iterative decoding. A wide spectrum of iterative decoders of varying complexity have been developed ranging from simple hard-decision algorithms such as Gallager-A/B algorithms to the more sophisticated belief propagation (BP) algorithm. Recently, the design of quantized BP decoders and other low-complexity variants of BP have gained prominence due to the high-speed requirements and hardware constraints for practical realizations of these decoders. The first quantized decoders including a three-level decoder coined Gallager-E algorithm were proposed by Richardson and Urbanke \cite{richardson}. They also developed the technique of density evolution to determine the asymptotic decoding thresholds of a code. Low complexity approximations to BP  with minimal loss in the asymptotic decoding thresholds have been proposed by Chen {\it et al.} \cite{lowcomplexchen}.  The class of quantized BP decoders have been investigated by Lee and Thorpe \cite{leethorpe}. Quantized min-sum decoders have been proposed by Smith, Kschischang and Yu \cite{Kschischang}.

A common theme in all the aforementioned works is that the underlying basis for design of the quantized decoders is to maximize the decoding thresholds which holds only in the asymptotic case. Therefore, these quantization schemes do not guarantee a good performance on a practical finite-length code especially in the high signal-to-noise (SNR) region. In addition, the effects of quantization can also contribute to the error-floor phenomenon. Richardson introduced the notion of  {\it trapping sets}  in \cite{richardsonerrorfloor} to characterize error floors. Trapping sets can be present in any finite-length code irrespective of how good the decoding threshold is and hence, codes optimized for good decoding thresholds can still exhibit high error floors. Characterization of error floors and design of LDPC codes with low error floors has recently been a subject of wide interest \cite{shashierrorfloor,shashiITW,ywang,zhangerrorfloor}.

In this paper, we propose multilevel decoders for LDPC codes over the binary symmetric channel (BSC). A key distinction from the traditional quantized decoders is that the messages are not quantized values of beliefs and the update rules are not approximations of the rules used in BP. Instead, they are derived using trapping sets and trapping set ontology \cite{ontology}. As we showed in \cite{milos} in the case of BSC, failure characterization is combinatorial in nature, and in the error floor region reduces to the problem of guaranteed error-correction capability of a code. In \cite{lucille} we showed the potential of multilevel decoding for the case of four levels. In this paper, we provide two 3-bit decoders for 3-left-regular codes that outperform floating-point BP and min-sum in the error floor region inspite of having much lower complexity. The rest of the paper is organized as follows. Section \ref{sect_Prelim} provides preliminaries. Section \ref{sect_framework} provides the general framework of multilevel decoders. In Section \ref{sect_threebits}, we provide the description of the 3-bit decoders for 3-left-regular codes. Finally results and conclusions are presented in Sections \ref{sect_results}  and \ref{sect_conclusion}.


\section{Preliminaries}\label{sect_Prelim}
Let $G=(V\cup C,E)$ denote the Tanner graph of a binary LDPC code $\cal{C}$ with the set of variable nodes $V=\{v_1,\cdots,v_n\}$ and set of check nodes $C=\{c_1,\cdots,c_m\}$. $E$ is the set of edges in $G$. The code has length $n$ and code rate $R$. For a vector $\mathbf{v}=(v_1,v_2,\ldots,v_n)$, the support of $\mathbf{v}$ denoted as $\supp(\mathbf{v})$, is defined as the set of all variable nodes such that $v_i\neq 0$. A code $\cal{C}$ is said to be $d_v$-left-regular if all variable nodes in $V$ of graph $G$ have the same degree $d_v$. The degree of a node is the number of its neighbors.

Let $\mathbf{r}=(r_1,r_2\ldots,r_n)$  be the input to the decoder from the BSC. A trapping set $\mathbf{T}(\mathbf{r})$ is a non-empty set of variable nodes in $G$ that are not eventually corrected by the decoder \cite{richardsonerrorfloor}. A standard notation for a trapping set is $(a,b)$ where $a=|\mathbf{T}(\mathbf{r})|$ and  $b$ is the number of odd-degree check nodes in the sub-graph induced by $\mathbf{T}(\mathbf{r})$. The critical number of a trapping set is the minimal number of variable nodes that have to be initially in error for the decoder to end up in the trapping set. Note that these definitions are for the case of BSC. Also in analysis of decoders in this paper,  it is implicitly assumed that the all-zero codeword is transmitted. This is a valid assumption since we consider only symmetric decoders, as explained in \cite{richardson}. Readers can refer to the work of Chilappagari {\it et al.} \cite{shashierrorfloor,ontology} for more details on these notions.

\section{Multilevel decoders: General Framework}\label{sect_framework}
Multilevel decoders are a new class of quantized message-passing decoders for LDPC codes. For this class, a decoder $\mathscr{F}$ is defined as a 4-tuple given by $\mathscr{F}=(\mathcal{M},\mathcal{Y},\Phi_v,\Phi_c)$. The message set $\cal{M}$ consists of all the levels under which the messages are confined to and is defined as $\mathcal{M}=\{ 0,\pm L_i \ :1\leq i\leq M  \}$, where $L_i\in\mathbb{R^{+}}$ and $L_i>L_j$ for any $i>j$. The set $\mathcal{Y}$  denotes the set of possible values called {\it channel values}, that are computed by the decoder based on the vector $\mathbf{r}$ received from the channel. For the case of BSC, $\mathcal{Y}$ is defined as $\mathcal{Y}=\{\pm \mrm{C}\}$, and for each variable node $v_i$ in $G$, the channel value $y_i\in \cal{Y}$ is determined by $y_i=(-1)^{r_i}\mrm{C}$.
$\Phi_v:\mathcal{Y} \times \mathcal{M}^{d_v-1} \to \mathcal{M} $ denotes the update rule used at a variable node with degree $d_v$, which is a simple map derived using knowledge of trapping sets.
$\Phi_c: \mathcal{M}^{d_c-1} \to \mathcal{M} $ denote the check node map for a check node with degree $d_c$. We shall restrict ourselves to $d_v$-left-regular codes where the same map is used at every variable node. Also the update rules we consider are time-invariant rules, i.e., they do not change from iteration to iteration.

{\it Remark:} In this paper, by the term ``trapping sets",  we mean trapping sets that are known for message-passing decoders such as Gallager-A/B and/or BP decoders.

\subsection{Update rules}
In this paper, we consider the check node update function $\Phi_c$ to be 
\begin{displaymath}
\Phi_c(m_1,\ldots,m_{d_c-1}) = \left(\prod_{j=1}^{d_c-1}\sgn(m_j)\right) \min_{j \in \{1,\ldots,d_c-1\}}(|m_j|)
\end{displaymath}
where $\sgn$ denotes the standard signum function.

{\it Remark:} $\Phi_c$ defined in this manner corresponds exactly to the check node update rule of the min-sum algorithm.

Based on the definition of $\Phi_v$, we propose two subclasses of multilevel decoders: linear-threshold decoders and non-linear-threshold decoders.

{\it Linear-threshold (LT) decoders:} For these decoders, the function $\Phi_v$ determines its output by taking a sum of its inputs and comparing with a set of thresholds.  A threshold set $\mathscr{T}=\{T_1,T_2,\cdots,T_M\}$ where $T_i \in \mathbb{R^+}$, is defined in a way such that for any $T_p,T_q\in \mathscr{T}$, $T_p>T_q$ if $p>q$.  The function $\Phi_v$ is defined as
\begin{displaymath}
\Phi_v(m_1,m_2,\cdots,m_{d_v-1},y_i)=Q\left(\sum_{j=1}^{d_v-1}m_j+y_i \right)
\end{displaymath}
where $Q$ is the quantization function defined as
\begin{displaymath}
Q(x)=\Bigg\{ \begin{tabular}{cc}
 $L_i$, & $T_{i}\leq x < T_{i+1}$ \\
$-L_i$, & $-T_{i+1}<x \leq -T_i$ \\
0, & otherwise
\end{tabular}
\end{displaymath}
where $ i=1, 2,\ldots,M$ and $T_{M+1}=\infty$.

{\it Remark:} Note that although one might consider these decoders as special instances of a quantized min-sum decoder, the messages are still not beliefs. Also particular message sets and threshold sets for $\Phi_v$ are defined using knowledge of trapping sets which will be discussed later.

{\it Non-linear-threshold (NLT) decoders:} For these decoders, the function $\Phi_v$ determines its output by taking the sum of its incoming messages and a weighted channel value, and then comparing with a set of thresholds. The weight $\omega_c$ assigned to the channel value $y_i$ is computed using a symmetric non-linear function $\Omega:\mathcal{M}^{d_v-1}\to \{0,1\}$.
\begin{displaymath}
\Phi_v(m_1,m_2,\cdots,m_{d_v-1},y_i)=Q\left(\sum_{j=1}^{d_v-1}m_j+ \omega_c \cdot y_i \right).
\end{displaymath}

{\it Remark:} Due to the nonlinearity, $\Phi_v$ can output different outgoing messages for any two distinct sets of incoming messages even though the sum of them is the same. Hence these decoders are different from quantized min-sum decoders or any other existing quantized message-passing decoders.

It is evident from the expressions for $\Phi_v$ and $\Phi_c$, that the rules are symmetric for both LT and NLT decoders.

\subsection{Isolation assumption}
We now introduce a key concept called {\it isolation assumption} that enables us to analyze decoding on isolated subgraphs induced by trapping sets of a code. Analyzing decoders on isolated subgraphs is a crucial strategy required for deriving update rules. It is important to note that this is different from the independence assumption considered by Gallager \cite{gallager}. Before we delve into this, we first provide some motivation for the need of defining such a concept. 

The design of multilevel decoders is based on the knowledge of trapping sets for existing decoders. Given a list of trapping sets, the aim is to design a variable node update rule that guarantees correction of all of them. Consider the subgraph of a trapping set contained in the Tanner graph with some nodes initially in error. In order to verify whether a given rule succeeds on a trapping set, it is necessary not only to know the subgraph induced by the trapping set, but also the neighborhood of the induced subgraph and messages coming from this neighborhood. Since this neighborhood can be different for different Tanner graphs, the design of multilevel decoders becomes very complex. Therefore, to facilitate the design process, we work under the assumption that the messages to the nodes of the trapping set from the nodes outside the induced subgraph of the trapping set are known. We call this assumption the \textit{isolation assumption} to signify the fact that the trapping set can be considered in isolation from the rest of the graph and can be analyzed as an independent entity.  

Before we formally introduce this, we require the notion of computation tree and introduce some more notations.
\begin{defn}
\cite{frey} A {\it computation tree} corresponding to a message-passing decoder of the Tanner graph $G$ is a tree that is constructed by choosing an arbitrary variable node in $G$ as its root and then recursively adding edges and leaf nodes to the tree that correspond to the messages passed in the decoder up to a certain number of iterations. For each vertex that is added to the tree, the corresponding node update function in $G$ is also copied.
\end{defn}

Let $G$ be the Tanner graph of a 3-left-regular code. Let $H$ be the induced subgraph of a trapping set $(a,b)$ contained in $G$ with variable node set $P\subseteq V$ and check node set $W\subseteq C$. Let $\mathcal{N}(u)$ denote the set of neighbors of a node $u$. Let $\mathcal{T}_{i}^k(G)$ be the computation tree of graph $G$ corresponding to a decoder $\mathscr{F}$ enumerated for $k$ iterations with variable node $v_i\in V$ as its root. Let $W^{\prime}\subseteq W$ denote the set of degree-one check nodes in the subgraph $H$. Let $P^{\prime}\subseteq P$ denote the set of variable nodes in $H$ where each variable node has at least one neighbor in $W^{\prime}$. During decoding on $G$, for a node $v_i\in P^{\prime}$,  let $\mu_l$ denote the message that $v_i$ receives from its neighboring degree-one check node in $H$ in the $l^{th}$ iteration.
\begin{defn}
A vertex $w\in \mathcal{T}_{i}^k(G)$ is said to be a {\it descendant} of a vertex $u\in \mathcal{T}_{i}^k(G)$ if there exists a path starting from vertex $w$ to the root $v_i$ that traverses through vertex $u$. The set of all descendants of the vertex $u$ in $\mathcal{T}_{i}^k(G)$ is denoted as $\mathcal{D} (u)$. For a given vertex set $U$, $\mathcal{D} (U)$ (with some abuse of notation) denotes the set of descendants of all $u\in U$.
\end{defn}

\begin{defn}
$\mathcal{T}_{i}^k(H)$ is called the computation tree of the subgraph $H$ enumerated for $k$ iterations for the decoder $\mathscr{F}$, if $\forall \ c_j\in W^{\prime}$, $\mu_l$ is given for all $l\leq k$, and if the root node $v_i \in P$ requires only the messages computed by the nodes in $H$ and $\mu_l$ to compute its binary hard-decision value.
\end{defn}
%

\begin{defn}
[Isolation assumption] The computation tree $\mathcal{T}_i^k(G)$ with the root $v_i\in P$ is said to be isolated if: (i) for any check node $c_j\in W^{\prime}$ that is in $\mathcal{T}_{i}^k(G)$ with $v_t\in P^{\prime}$ as its parent, $\mathcal{D} (c_j) \cap \mathcal{D} (\mathcal{N}(v_t)\setminus c_j) = \emptyset$, and (ii) for any two check nodes $c_r,c_s \in W\setminus W'$ that are also in $\mathcal{T}_{i}^k(G)$, $\mathcal{D} (c_r)\cap \mathcal{D} (c_s)\subseteq (P\cup W)$. If $\mathcal{T}_i^k(G)$ is isolated $\forall v_i \in P$, then the subgraph $H$ is said to satisfy the isolation assumption in $G$ for $k$ iterations.
\end{defn}

{\it Remark:} The isolation assumption can still be satisfied even when there are nodes in $H$ that appear multiple times in $\mathcal{T}_{i}^k(G)$ as long as these nodes are not descendants of the degree-one check nodes. Whereas Gallager's independence assumption will be violated if any node in $H$ is repeated in $\mathcal{T}_{i}^k(G)$. Hence, isolation assumption is a weaker condition than independence. For clarity, we illustrate with an example shown in Fig. \ref{graph62}.

\begin{ex}
Let us assume that the graph $G$ of code $\mathcal{C}$ contains a subgraph $H$ induced by a $(6,2)$ trapping set. Fig. \ref{graph62} shows the subgraph $H$, and the computation tree $\mathcal{T}_{3}^2(G)$ of graph $G$ with $v_3$ as its root enumerated for two iterations. The {\scriptsize${\blacksquare}$} denotes a odd-degree check node. The solid lines represent connections within subgraph $H$ and the dotted lines represent connections from the rest of the graph $G$ outside the subgraph $H$. The isolation assumption is satisfied for two iterations if none of the descendants of the check nodes $c_7$ and $c_8$  appear as a descendant of check node $c_9$ (similar condition has to hold for $c_{10}$), and if the only common descendants of the degree-2 check nodes are nodes in $H$. But the independence assumption does not hold for two iterations.
\begin{figure}
\begin{center}
\subfigure[] 
{
\label{6_2Trap}
\includegraphics[angle = 0,width=1.6in]{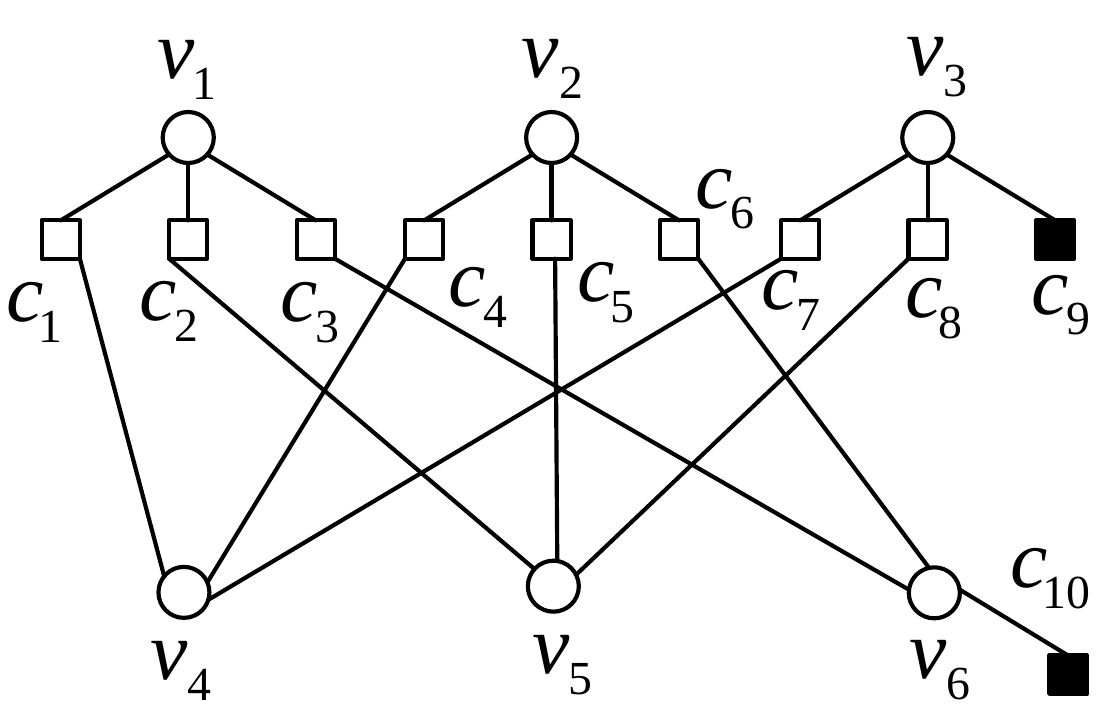}
}
\subfigure[] 
{
\label{comptree}
\includegraphics[angle = 0,width=1.6in]{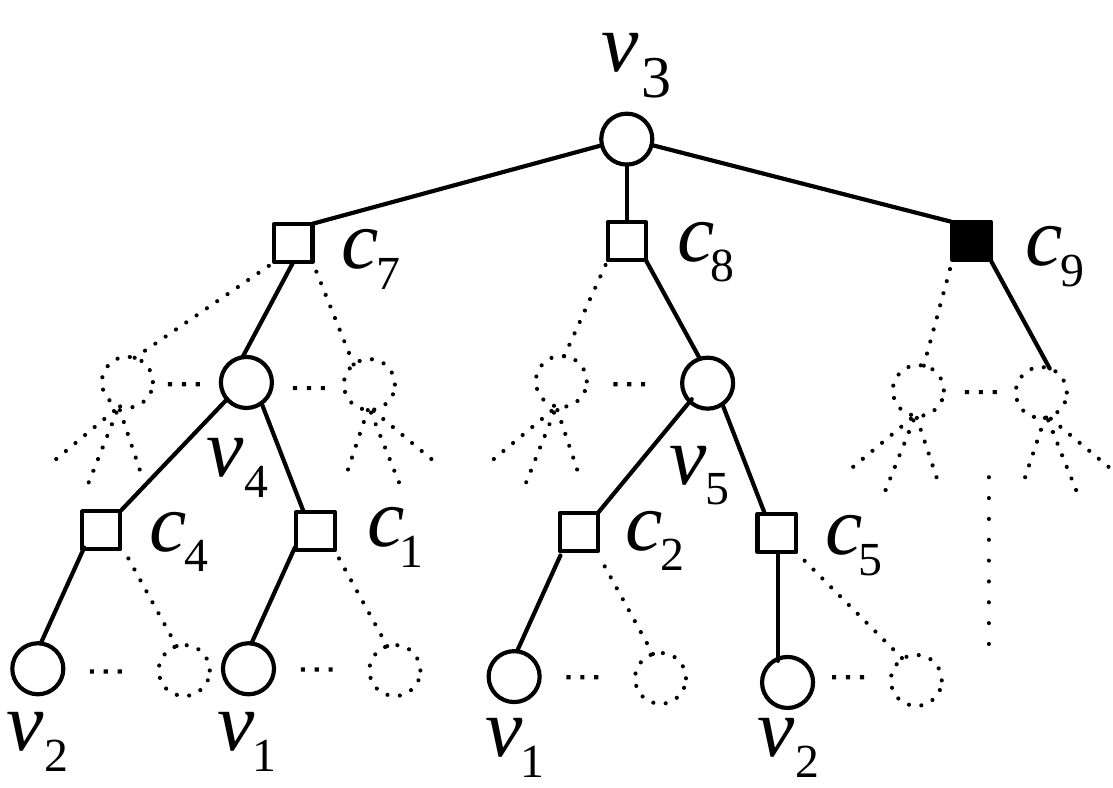}
}
\end{center}
\caption{Subgraph $H$ induced by (6,2) trapping set contained in $G$: \subref{6_2Trap} Tanner graph of $H$; \subref{comptree} computational tree $\mathcal{T}_{3}^2(G)$}
\label{graph62}
\vspace{-0.2in}
\end{figure}
\end{ex}

\begin{te}
[Isolation theorem] Let $G=\{V\cup C,E\}$ of a 3-left-regular code which contains a subgraph $H=\{P\cup W,E'\}$ that is induced by a trapping set $(a,b)$. Let $W^{\prime}\subseteq W$ denote the set of degree-one check nodes in $H$ and let $P^{\prime}\subseteq P$ denote the set of variable nodes in $H$  where each has at least one neighbor in $W^{\prime}$. If $\mathbf{r}$ is input to decoder $\mathscr{F}$ from the BSC such that $\supp(\mathbf{r})\in P$, and if $H$ satisfies the isolation assumption in $G$ for $k$ iterations,  then for each $c_j\in W^{\prime}$, the message from $c_j$ to its neighbor in $H$ in the $l^{th}$ iteration denoted by $\mu_l$, is determined as the output of $\Phi_v(\mu_{l-1}, \mu_{l-1},\mrm{C})$ $\forall l\leq k$.
\end{te}
\IEEEproof
This follows by looking at the computation tree $\mathcal{T}_{i}^l(G)$ where $l\leq k$ with any $v_i\in P^{\prime}$ as its root. Let the initial messages passed from a variable node be $\pm \mu_0\in \mathcal{M}$. Since $\supp(\mathbf{r})\in P$, due to the isolation assumption, this means that all the variable nodes that are descendants to any $c_j\in W^{\prime}$ are initially correct. In the initial iteration, from the definition of $\Phi_c$, the outgoing message of check nodes that are descendants to $c_j$ is $\mu_0$. In the next iteration, the variable nodes connected to these check nodes receive $\mu_0$ on all their edges due to the isolation assumption and send $\Phi_v(\mu_0,\mu_0, \mrm{C})$ as their outgoing messages. Due to the definition of $\Phi_c$, the check nodes connected to these nodes in $\mathcal{T}_{i}^l(G)$ send $\mu_1$ which is simply $\Phi_v(\mu_0,\mu_0, \mrm{C})$. This process inductively follows while traversing up the tree for $l$ iterations. Moreover, computation of the hard-decision value for any node $v_i$ in $H$ requires only messages from nodes in $H$ in addition to $\mu_l$.
\endIEEEproof
{\it Remark:} Note that the isolation assumption and theorem can be restated for the min-sum decoder.

\begin{cor}
Consider the min-sum decoder for 3-left-regular LDPC codes with $\mathcal{Y}=\{\pm 1\}$. If subgraph $H$ contained in $G$ satisfies the isolation assumption for $k$ iterations, and if all variable nodes outside $H$ are initially correct, then $\mu_l$ of the degree-one check node for the min-sum decoder is $2\mu_{l-1} + 1$.
\end{cor}
\begin{cor}
If $H$ is a subgraph contained in $G$ such that it satisfies the isolation assumption for $k$ iterations, and if all variable nodes outside $H$ are initially correct, then the computation tree $\mathcal{T}_{i}^k(G)$ with $v_i\in P$  is equivalent to $\mathcal{T}_{i}^k(H)$, provided $\mu_l$ for each degree-one check node in $H$ is computed using the isolation theorem.
\end{cor}

{\it Remark:} The above corollary validates decoding on isolated subgraphs that are induced by trapping sets and this property is useful for deriving good update rules. Note that the number of iterations required for a subgraph to satisfy the isolation assumption in order to converge becomes an important parameter. Also the notion of critical number can now  be extended for multilevel decoders as well.

\section{3-bit decoders for 3-left-regular LDPC codes}\label{sect_threebits}
We provide two particularly good 3-bit decoders: a 7-level LT decoder, and a 5-level NLT decoder for 3-left-regular LDPC codes. These were derived by considering a systemtatic hierarchy of trapping sets called trapping set ontology \cite{ontology}, and using the approach described in the previous section. Some important criteria to be considered in the derivation of the rules are increase in critical number and convergence in fewer iterations. Due to space constraints, we do not give details of deriving good rules in this paper but we shall demonstrate how multilevel decoders can correct certain error patterns uncorrectable by even floating-point algorithms.

The function $\Phi_v$ can be uniquely defined by setting constraints on the magnitudes and thresholds. We shall use this approach to define the decoders.

For the 7-level LT decoder, the constraints that uniquely define the decoder are  $L_1<C<2L_1$, $L_2=2L_1+C$, $L_3=2L_2 + C$, and $T_1=L_1$, $T_2=L_2$, $T_3=L_3$.

For the 5-level NLT decoder, the constraints that specify the decoder are $C=L_1$, $L_2=3L_1$, $T_1=L_1$, $T_2=L_2$, and the channel weight function used to compute $\omega_c$ is given by
\begin{eqnarray}
\omega_c &=&\Omega(m_1,m_2) \nonumber \\
&=& 1-\big(\mrm{sign}(m_1) \oplus\mrm{sign}(m_2) \big) \cdot \delta(|m_1|+|m_2|- 2 L_2 ).  \nonumber
\end{eqnarray}

As an example, we now illustrate how a 3-error pattern on a $n=786$, $R=0.75$ quasicyclic code \cite{website} that was uncorrectable by min-sum decoder, is correctable by the 7-level LT decoder. 
\begin{ex}\label{ex95}
Let $H$ be the subgraph induced by the (9,5) trapping set which contains a (6,2) and has three degree-one checks as shown in Fig. \ref{graph95}. Consider the 3-error pattern shown in the figure where {\large$\bullet$} denotes an initially wrong variable node.
\begin{figure}[bhtp]
\begin{center}
\includegraphics[angle=0, width=2.4in]{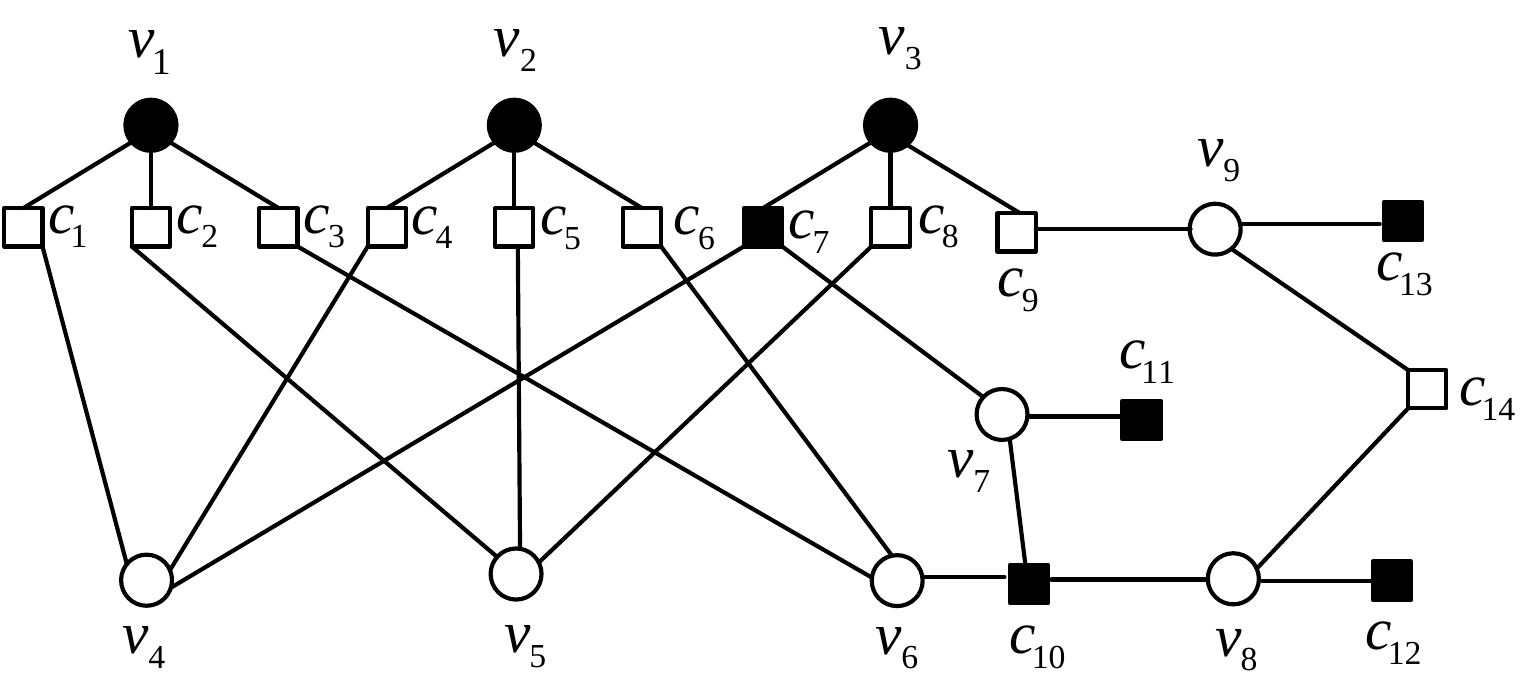}
\caption{Subgraph induced by (9,5) trapping set that contains a (6,2)}
\label{graph95}
\end{center}
\end{figure}

Under the isolation assumption, let us analyze the decoding of 7-level LT decoder on the subgraph with the help of the isolation theorem.
Let $m_k(v_i,:)$ denote all outgoing messages of node $v_i$ in the first half of $k^{th}$ iteration and let $m_k(:,v_i)$ denote all incoming messages to node $v_i$ from checks in the second half of $k^{th}$ iteration. We will show only messages passed by certain crucial nodes in each iteration. Assume that all messages are initially zero.

In first half of iteration 1, all outgoing messages are $\pm L_1$, i,e, $m_1(v_i,:)=(-L_1,-L_1,-L_1)$ for $i\in \{1,2,3\}$ and $m_1(v_i,:)=(L_1,L_1,L_1)$ for $ i\notin \{1,2,3\}$. In the second half, check nodes send their messages by the isolation assumption. Then $m_1(:,v_i)=(-L_1,-L_1,-L_1)$ for $i\in \{4,5\}$.
%

In the first half of iteration 2, because $\Phi_v(L_1,L_1,\mrm{-C})=0$, and $\Phi_v(-L_1,-L_1,\mrm{C})=0$, this update rule helps prevent nodes $v_4$ and $v_5$ from sending wrong messages. Then $m_2(v_i,:)=0$ for $i\in \{1 \ \mrm{to} \ 5\}$. Check nodes send their messages in the second half and $m_2(:,v_i)=(0,0,L_1)$ for $i\in \{1,2\}$.

Finally in the first half of iteration 3, nodes $v_1$ and $v_2$ are the only nodes that can send wrong messages. But because $\Phi_v(0,L_1,-\mrm{C})=0$, the nodes send zero instead, and the decoder converges at the end of iteration 3.
\end{ex}

{\it Remark:} From the above example, we see that certain outputs of $\Phi_v$ were crucial for preventing propagation of wrong messages and convergence within 3 iterations. Whereas the min-sum decoder requires 4 iterations under the isolation assumption to converge on the same 3-error pattern. Now if subgraph $H$ contained in $G$ satisfies the isolation assumption for only 3 iterations, then min-sum is not guaranteed to correct three errors. For the quasicyclic code, this particular 3-error pattern on such a subgraph fails to be corrected by min-sum but is corrected by 7-level LT decoder. In fact, 7-level LT decoder corrects all 3-error patterns on the code. 

Although we considered error patterns that failed to decode by the min-sum in example \ref{ex95}, the same analysis can be carried out on error patterns that failed to decode by BP as well. The rules can be derived to correct such patterns in a similar fashion by ensuring convergence in fewest number of iterations under the isolation assumption. For example, the 7-level LT decoder did not fail for any 4-error patterns on the same quasicyclic code whereas the BP and min-sum decoders failed in the region of simulation in Fig. \ref{Quasi}.

\section{Numerical Results}\label{sect_results}
\begin{figure}
\begin{center}
\includegraphics[angle=0, width=3in]{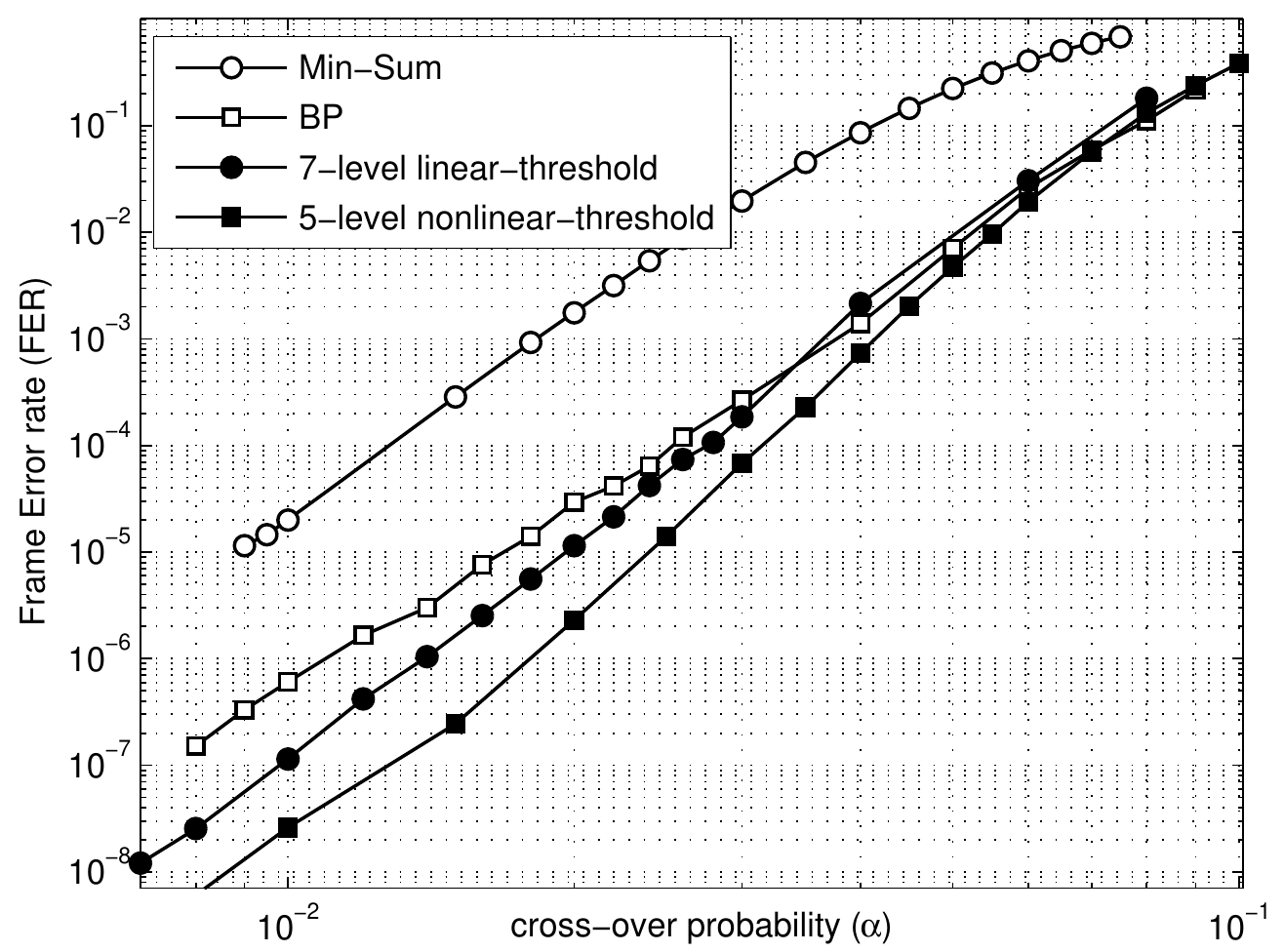}
\caption{FER results on the $n=155$, $R=0.4$, Tanner code}
\label{Tanner}

\end{center}
\end{figure}

\begin{figure}
\begin{center}
\vspace{-0.15in}
\includegraphics[angle=0, width=3in]{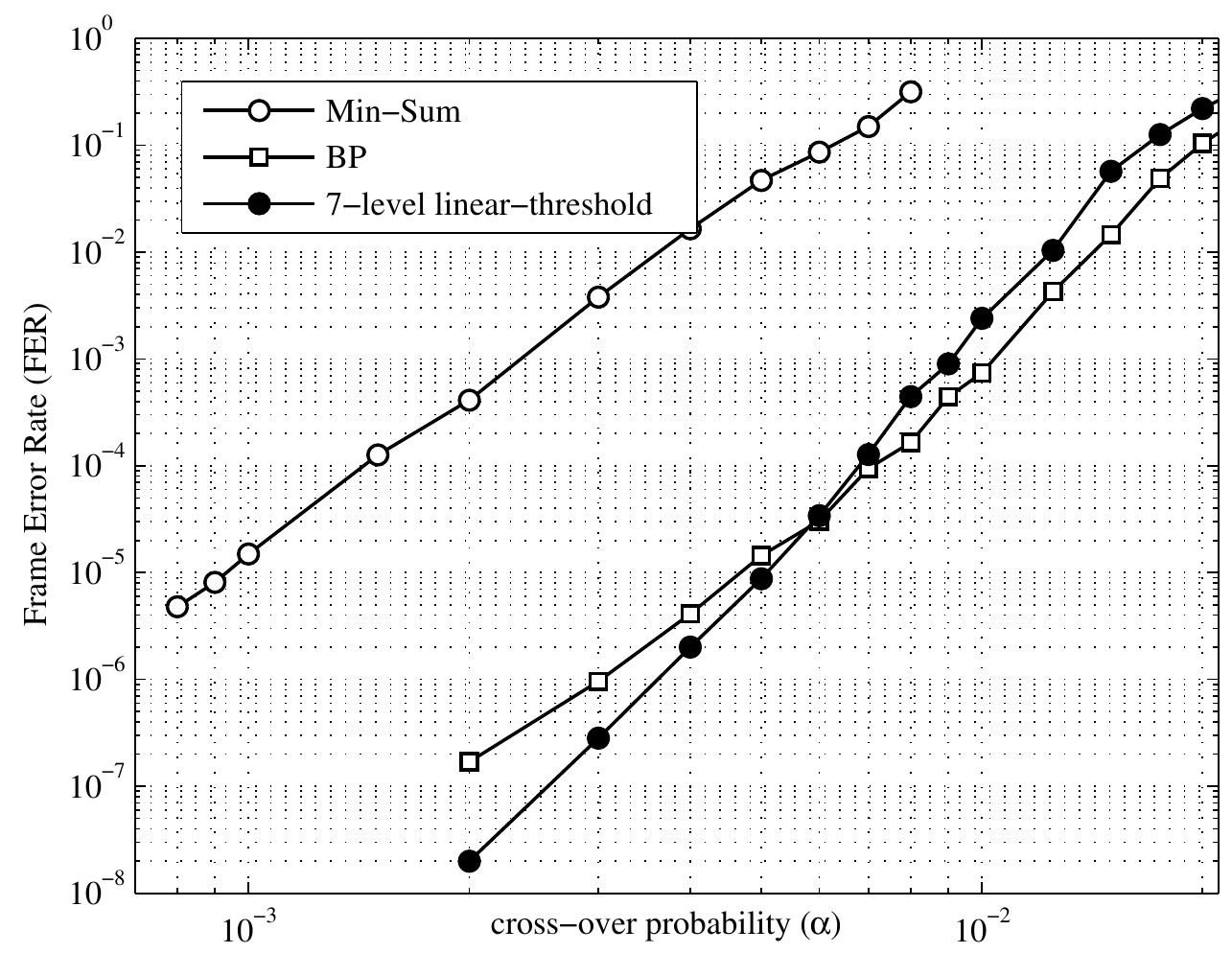}
\caption{FER results on the $n=768$, $R=0.75$, Quasicyclic code}
\vspace{-0.2in}
\label{Quasi}
\end{center}
\end{figure}

\begin{figure}
\begin{center}
\includegraphics[angle=0, width=3in]{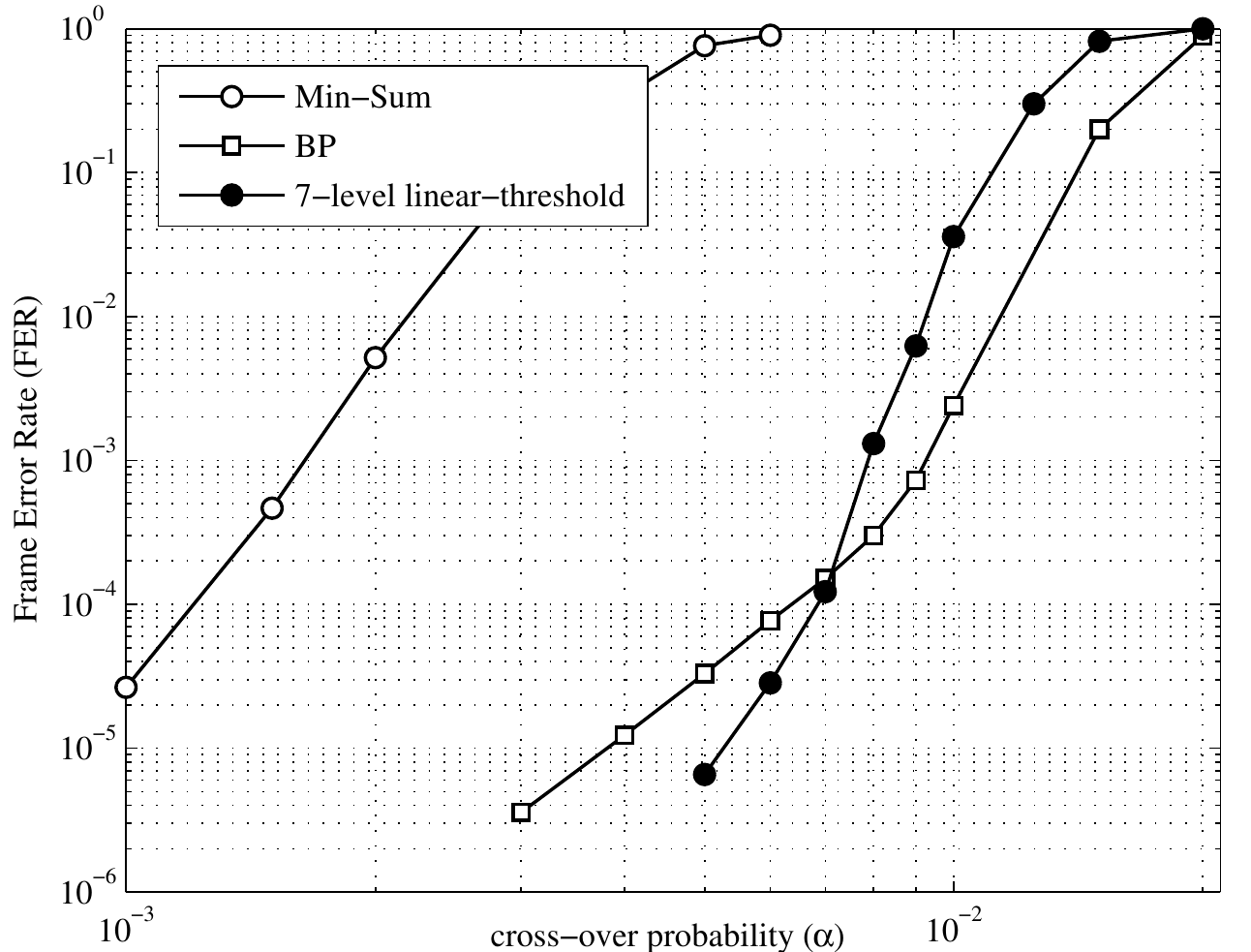}
\caption{FER results on the $n=4085$, $R=0.82$, MacKay code}
\label{MacKay}
\end{center}
\vspace{-0.1in}
\end{figure}

Simulations for frame error rate (FER) were carried out on three different codes: 1) $n=155$, $R=0.4$, Tanner code, 2) $n=768$, $R=0.75$, Quasicyclic code with $d_{min}=12$, and 3) a $n=4085$, $R=0.82$, MacKay code. The codes were chosen to cover a broad spectrum of LDPC codes in order to validate our approach. The Tanner code is well-understood and has been analyzed for many different decoders. The high-rate quasicyclic code was chosen since the error floor problem is much more challenging for high-rate codes. A MacKay code was chosen as an example of a high-rate random code. Fig. \ref{Tanner}, Fig. \ref{Quasi}, and Fig. \ref{MacKay} show the simulation results. The maximum number of iterations used was 100 for all decoders. Structures of these three codes can be found in \cite{website}.

For all three codes, the 3-bit decoders significantly outperform BP in the error floor region. Notice the difference in slopes in the FER curves. For the Tanner code, the 5-NLT decoder guarantees correction of all error patterns up to 5 errors. An interesting point to note is that the highly complex linear programming (LP) decoding fails to correct all 5-errors on the Tanner code, whereas the 3-bit decoder is able to, which illustrates the power of multilevel decoding. 


%

\section{Discussion and Conclusions}\label{sect_conclusion}
Multilevel decoding was established as a powerful decoding technique for LDPC codes. From our concluding example, we highlighted the importance of the isolation assumption as a crucial strategy for deriving good rules. Also from the results, it is evident that there is a certain aspect of universality to these decoders; the same 3-bit decoder performs well on a variety of codes.  This suggests that deriving good rules based on trapping sets appears to take the structure of the local neighborhood into local computations of messages, which is a tiny step closer in our pursuit to approach maximum-likelihood decoding. Our future work includes deriving bounds and relating the isolation theorem with guaranteed error-correction capability.

%
\section*{Acknowledgment}
This work was funded by NSF under the grants IHCS-0725403, CCF-0634969, CCF-0830245.

\end{document}